\documentclass[aip,reprint,notitlepage,onecolumn,nofootinbib]{revtex4-2}
\usepackage{bm,graphicx,xcolor,hyperref,titlesec,multirow,commath,enumitem,orcidlink}
\hypersetup{colorlinks,citecolor=blue,linkcolor=red,urlcolor=blue}
\setcitestyle{numbers,square,sort&compress}
\usepackage[utf8]{inputenc}
\usepackage[T1]{fontenc}
\titleformat{\section}{\large\sffamily\bfseries}{\thesection}{1em}{}
\titleformat{\subsection}{\normalsize\sffamily\bfseries}{\thesubsection}{1em}{}
\titlespacing\section{0pt}{12pt plus 4pt minus 2pt}{8pt plus 2pt minus 2pt}
\titlespacing\subsection{0pt}{10pt plus 4pt minus 2pt}{6pt plus 2pt minus 2pt}
\titlespacing\subsubsection{0pt}{10pt plus 4pt minus 2pt}{6pt plus 2pt minus 2pt}

\renewcommand{\thesection}{\arabic{section}}
\renewcommand{\thesubsection}{\thesection.\arabic{subsection}}

\makeatletter
\renewcommand{\p@subsection}{}
\renewcommand{\p@subsubsection}{}
\makeatother
\newcommand*{\xdash}[1][3em]{\medskip\rule[1ex]{#1}{0.5pt}}

\begin{document}
\title{Thermodynamics-Like Formalism for Immiscible and Incompressible
  Two-Phase Flow in Porous Media}

\author{Alex Hansen\,\orcidlink{0000-0002-0860-3880}}
\email{Alex.Hansen@ntnu.no}

\author{Santanu Sinha\,\orcidlink{0000-0003-0838-3096}}
\email{Santanu.Sinha@ntnu.no}
\affiliation{PoreLab, Department of Physics, Norwegian University of
  Science and Technology NTNU, N-7491 Trondheim, Norway\looseness=-1}

\date{\today}

%%%%%%%%%%%%%%%%%%%%%%%%%%%%%%%%%%%%%%%%%%
\begin{abstract}
  It is possible to formulate immiscible and incompressible
  two-phase flow in porous media in a mathematical framework
  resembling thermodynamics based on the Jaynes generalization of
  statistical mechanics. We review this approach and discuss the
  meaning of the emergent variables that appear, agiture, flow
  derivative and flow pressure, which are conjugate to the
  configurational entropy, the saturation and the porosity
  respectively.  We conjecture that the agiture, the temperature-like
  variable, is directly related to the pressure gradient. This has as
  a consequence that the configurational entropy, a measure of how the
  fluids are distributed within the porous media and the accompanying
  velocity field, and the differential mobility of the fluids are
  related. We also develop elements of another version of the
  thermodynamics-like formalism where fractional flow rather than
  saturation is the control variable, since this is typically the
  natural control variable in experiments.
\end{abstract}
%%%%%%%%%%%%%%%%%%%%%%%%%%%%%%%%%%%%%%%%%%
\keywords{Flow in porous media; Immiscible two-phase flow; Co-moving velocity} 
%%%%%%%%%%%%%%%%%%%%%%%%%%%%%%%%%%%%%%%%%%

\maketitle

\section{Introduction}
\label{intro}

In spite of immisicble multi-phase flow in porous media being a field
of great industrial importance for more than a century, basic research
in this field has not received the attention that it deserves. A main
reason for this is that the field is fragmented, being spread over
many different disciplines such as biology, geology, geophysics,
physics, chemistry, and materials science.  Another reason is that it
is easy to get the impression that it is not ``clean" in the sense
that it is a field where theories may be built.  Rather it seems to be
a collection of phenomena in need of classification.  It is the aim of
this paper to convince the readership that this is indeed not the
case.  There is a physics of flow in porous media \cite{ffh22} where
general principles may be formulated.  We will in the following
present the readers with a review of our ongoing attempts at
formulating a mathematical framework for immiscible and incompressible
two-phase flow in porous media.

In 1907, Buckingham introduced several of the central concepts used
today in describing immiscible and incompressible fluid flow in porous
media \cite{b07,n07}, but without joining them through equations. This
was first done by Richards in 1931 \cite{r31}, who --- as Buckingham
--- placed the flow problem in the context of water movement in soils.
In 1936, building on the work by Wyckoff and Botset \cite{wb36},
Muscat and Meres \cite{mm36} introduced the concept of relative
permeability, a concept generalizing the notion of ``capillary
conductance" that Buckingham introduced \cite{b07}.  With the
introduction of the capillary pressure curve by Leverett \cite{l40}
into the framework established by Muscat and Meres, the generalized
Darcy constitutive equations that are generally used today in
e.g.\ reservoir simulation, were in place.  The formulation of the
problem in the context of water in soils that Buckingham pioneered and
which was followed by the work of Richards forms an alternate
formulation of essentially the same ideas.

We review briefly relative permeability theory --- the only approach
to immiscible two-phase flow in porous media in practical use today
--- in Section \ref{relperm}.  Relative permeability theory is purely
phenomenological and has well-known deficiencies. As a result, there
is an ongoing effort to derive effective equations going beyond
relative permeability theory based on the physics at the pore scale.
This effort is based on {\it homogenization\/} which we review briefly
in Section \ref{homogenization}.

In Section \ref{statmech}, we introduce our approach to immiscible
two-phase flow in porous media based on a generalization of
statistical mechanics by Jaynes \cite{j57,hfss23}. Statistical
mechanics in its traditional role is to connect a molecular
description of matter to a description at the much larger scales where
matter appears continuous, namely thermodynamics.  In adopting this
idea to two-phase flow in porous media, we find a thermodynamics-like
framework on large scales that together with conservation laws provide
a closed set of equations that has the equations of relative
permeability theory as a special case.  As in ordinary thermodynamics,
the difference between intensive and extensive variables is
central. We explore this in Section \ref{intensive}. The intensive
variables appearing in the thermodynamics-like formalism we derive are
emergent in that they have no equivalent on the pore scale.  This is
discussed in Section \ref{emergent}.  The temperature-like intensive
variable, {\it agiture,\/} we conjecture to be proportional to the
pressure gradient in Subsection \ref{agiture}. This conjecture has as
a consequence that the {\it configurational entropy,\/} which is an
example of Shannon's information entropy, is related to the
differential mobility\footnote{The differential mobility is the
derivative of the flow velocity with respect to the pressure gradient.
The mobility is the flow velocity divided by the pressure gradient.}
of the fluids. The {\it flow derivative\/} we discuss in Subsection
\ref{mu} is the conjugate variable to the saturation which is the
fractional volume. We end the section by discussing the conjugate of
the porosity, the {\it flow pressure\/} in Subsection \ref{pi}.

In Section \ref{seepage}, we introduce the {\it co-moving velocity\/}
which relates the thermodynamic velocities --- corresponding to
partial molar volumes in ordinary thermodynamics --- to the seepage
velocities which are the average pore velocity of each of the
immiscible fluids.  This concept has the potential to become a very
useful concept in practical calculations.

Section \ref{control} contains material that has not been presented
before. We start by pointing out a choice made when constructing the
Jaynes statistical mechanics in Section \ref{statmech}, and how it
could have been made differently.  Then, in Subsection \ref{flowrate},
we discuss what the ensuing thermodynamics-like formalism would be as
a consequence: rather than having saturation as one of the control
parameters, we find fractional flow rate being a control parameter.
This is useful as this is the control parameter used in core flooding
experiments.

We end by summarizing and concluding in Section \ref{discussion}.

%%%%%%%%%%%%%%%%%%%%%%%%%%%%%%%%%%%%%%%%%%
\section{Relative permeability theory}
\label{relperm}

The core idea of the relative permeability approach may be summarized
as follows: We have two immiscible fluids, one more wetting with
respect to the porous matrix than the other fluid.  We will refer to
them as the wetting (w) and non-wetting (n) fluid respectively.  From
the perspective of the wetting fluid, the pore space it sees is the
total pore space of the porous medium minus the pore space occupied by
the non-wetting fluid, and vice versa.  The relative reduction of pore
space for each fluid implies a reduction in effective permeability for
each fluid.  We express these statements in the form of two
constitutive equations, one for each fluid \cite{ffh22},
\begin{eqnarray}
{\vec v}_w&=&-\ \frac{Kk_{rw}(S_w)}{\mu_w \phi S_w} \nabla P_w\;,\label{eq1-1}\\
{\vec v}_n&=&-\ \frac{Kk_{rn}(S_w)}{\mu_n \phi S_n} \nabla P_n\;.\label{eq1-2}
\end{eqnarray}
Here ${\vec v}_w$ and ${\vec v}_n$ are the pore velocities of the
wetting and non-wetting fluids, $K$ is the (isotropic) permeability of
the porous medium, $\mu_w$ and $\mu_n$ are the viscosities of each
fluid, $\phi$ is the porosity of the porous medium, $P_w$ and $P_n$
are the fluid pressures, $S_w$ and $S_n$ are the wetting and
non-wetting saturations, and $k_{rw}(S_w)$ and $k_{rn}(S_w)$ are the
relative permeabilities of the two fluids. It is an essential {\it
  assumption\/} in this theory that these two quantities are functions
of the saturations alone. The saturations $S_w$ and $S_n$ are defined
as the fraction of pore space occupied by each fluid so that
\begin{equation}
\label{eq1-3}
S_w+S_n=1\;.
\end{equation}
The difference in pressure between the two fluids is defined as the
capillary pressure curve $P_c$,
\begin{equation}
\label{eq1-4}
P_n-P_w=P_c(S_w)\;.
\end{equation}
It is an assumption that the capillary pressure curve only depends on
the saturation $S_w$. The capillary pressure curve is a particularly
difficult quantity, both conceptually and in terms of measurement, see
e.g.\ \cite{mfmst15}.  We define the average pore velocity ${\vec
  v}_p$ as
\begin{equation}
\label{eq1-5}
{\vec v}_p=S_w{\vec v}_w+S_n {\vec v}_n\;,
\end{equation}
i.e., we are using a volume average. This makes sense since we are
assuming the fluids to be incompressible:
\begin{equation}
\label{eq1-6}
\nabla \cdot [\phi {\vec v}_p]=0\;.
\end{equation}
Volume conservation then gives
\begin{eqnarray}
\phi \frac{\partial S_w}{\partial t} +\nabla\cdot \left[\phi S_w {\vec v}_w\right]&=&0\;,\label{eq1-7}\\   
\phi \frac{\partial S_n}{\partial t} +\nabla\cdot \left[\phi S_n {\vec v}_n\right]&=&0\;,\label{eq1-8}
\end{eqnarray}
where $t$ is time. This set of equations is closed as long as
$k_{rw}(S_w)$, $k_{rn}(S_w)$ and $P_c(S_w)$ are provided. In an
industrial context, Special Core Analysis --- SCAL --- forms the work
flow for obtaining this information.

%%%%%%%%%%%%%%%%%%%%%%%%%%%%%%%%%%%%%%%%%%
\section{Homogenization}
\label{homogenization}

It is well known that assuming the three constitutive functions depend
only on the saturation is generally not correct.  The relative
permeabilities do depend not only on the saturation, but also how the
fluids arrange themselves in the porous medium, and this depends in
turn on the flow rate.  The capillary pressure curve is highly
hysteretic, signaling that there are missing variables. However, it
has turned out to be difficult to go beyond this almost 90 years old
phenomenological framework.  The dominating approach is based on
homogenization.  This approach may be divided into two branches: That
which focuses on momentum transfer and that which focuses on energy
transfer.  In the first case, one starts with the hydrodynamic
equations on the pore scale \cite{as86,a87,alb89,pb18,pb19,lv22}, and
in the second case, one sets up a thermodynamic description at the
pore scale \cite{hg79,hg90,hg93,hg93b}. Then comes the
coarse-graining, i.e., spatial averaging.  There are several ways of
doing this, perhaps that of Whitaker is the most well-known
\cite{w86,w86a}.  It is based on equating the average of the gradient
of a variable associated with pore space to the gradient of the
average variable plus an integral over the surface area of the pores.
Since the surface area of porous media scales as the volume (i.e., the
surface area is {\it extensive\/} --- the defining property of porous
media), this integral does not vanish as one goes up in scale. We
split the variable appearing in the surface integral into an average
part and a fluctuating part.  This leaves us with having the average,
and gradients of the average expressed in terms of the fluctuations of
the original variable.  The last step is to make an independent
assumption on how the fluctuating variables is related to the averages
–-- a closure assumption.  With this, the equations between the
original, pore-scale variables have been turned into equations between
spatial averages of these variables.

Whereas the first of these two approaches is based on momentum
transfer, the second approach is based on energy transfer, i.e.,
thermodynamics. One starts with a thermodynamic description on the
pore level which is then homogenized.  This approach has developed
into Thermodynamically Constrained Averaging Theory (TCAT)
\cite{gm05,gm05a,nbh11,gm14}.

A recent paper by McClure et al.\ \cite{mfbablr22} attempts to derive
the relative permeability equations from an energy budget based on
thermodynamic considerations and homogenization.  The relative
permeability equations do appear as a first term in a series
expansion.  It is not shown in \cite{mfbablr22}, however, that the
higher order terms are negligible.

The topology of a geometric object such as a porous medium may be
described using the four {\it Minkowski functionals\/} volume, surface
area, mean curvature, and the Euler characteristics.  The {\it
  Hadwiger theorem\/} states that the Minkowski functionals form a
complete basis set for all extensive functions that are invariant with
respect to the orientation of the object \cite{m00}.  The use of this
theorem to characterize the free energy of fluids in a porous medium
combined with homogenization forms another approach to the scale-up
problem \cite{cacbsbgm18,kbsbt18,acbrlasb19,cba19}.  Homogenization as
seen so far is based on {\it spatial\/} averages only. McClure et
al.\ \cite{mab21,mba21} emphasize that there is also time which one
should average over and that different processes work on different
time scales.

 An approach circumventing the complexities associated with
 homogenization is based on classical non-equilibrium thermodynamics
 \cite{gm84,kbjg17}, see \cite{kbhhg19,kbhhg19b,wgbkch20,bk22}.  By
 using the extensiveness of the internal energy of the fluids in the
 porous medium, the Euler theorem for homogeneous functions allows for
 defining thermodynamic variables such as pressure and chemical
 potentials on the Darcy scale.  Gradients in the intensive variables
 are introduced and the machinery of classical non-equilibrium
 thermodynamics \cite{gm84,kbjg17} is then set in motion. The
 underlying homogenization is somewhat hidden in this approach, but it
 is underlying the way a Representative Elementary Volume (REV) is
 defined and used.

The main practical difficulty with homogenization is the complexity of
the ensuing equations and the large number of variables that are
necessary. The root of this difficulty is that homogenization can only
produce averages over the original variables, and thus cannot produce
new types of variables which go hand in hand with capturing emergent
properties \cite{a72}.

%%%%%%%%%%%%%%%%%%%%%%%%%%%%%%%%%%%%%%%%%%
\section{Statistical mechanics}
\label{statmech}

The essence of these efforts has been to find a description of
immiscible and incompressible two-phase flow in porous media at scales
where the medium acts as a continuum.\footnote{Including
compressibility would necessitate switching from conserved fluid
volume to conserved fluid mass. It would also couple the theory to
ordinary thermodynamics through equations of state for the
compressible fluids. We wish to avoid these complications at the
present stage.}  This is an example of {\it upscaling.\/} The
situation is thus that we have a continuum scale phenomenological
theory --- relative permeability theory that is simple enough to be
used in practical situations, but which is approximate. Attempts at
going beyond this theory has, however, not led to practical
applications due to the complexity that arises.  However, if we look
beyond the problem we discuss here, there is an example of successful
upscaling that is very familiar: The scale-up from a molecular
description to thermodynamics based on statistical mechanics.  At
molecular scales, the position and momentum of each molecule are the
variables that describe a monoatomic gas.  This works even if we are
dealing with millions of molecules in a molecular dynamics simulation
\cite{fs23}.  However, it is rather of the order $10^{23}$ molecules
that are relevant in e.g.\ chemistry.  A molecular description at such
scales is useless.  Here {\it emergent variables\/} such as
temperature and pressure adequately describe the gas.  Statistical
mechanics converts the molecular description to thermodynamics which
not only contains emergent variables but also relations between them.
These variables and their relations have no meaning within a molecular
description.

In the mid-fifties, Jaynes generalized statistical mechanics from
being a theory for molecular matter to being based on information
theory \cite{j57}. This was done by replacing Boltzmann's statistical
interpretation of thermodynamic entropy by Shannon's generalized
information entropy \cite{s48}.  The information entropy is a
quantitative measure of what is {\it not\/} known about a
system. Suppose we have a stochastic process that can produce $N$
possible outcomes.  We number the outcomes from 1 to $N$, assuming a
probability $p_i$ for the $i$th outcome, $x_i$.  A quantitative
measure of our ignorance about this process must be independent of how
we group events together.  For example, if we consider composite
events such as $p_{ij}=p_i p_j$, i.e., the probability that an event
$i$ is followed by an event $j$, then the information entropy must
remain the same.  If we know nothing about the process, its
information entropy must be at a maximum, since our ignorance is
maximal.  In order to utilize this, Shannon generalized to $N$
outcomes the Laplace principle of insufficient reason \cite{l03} which
states that the optimal choice of probabilities for a stochastic
process with {\it two\/} possible outcomes is to assign them equal
probability. Hence, the optimal choice when there are $N$ outcomes is
to assign all probabilities the same value, $p_i=1/N$.  There is only
one function that can be constructed from $p_i$ with these properties,
\begin{equation}
\label{eq4-1}
S_I=-\sum_{i=1}^N p_i \ln p_i\;,
\end{equation}
which is then the information or Shannon entropy. Jaynes then posed
the question, what happens if we {\it do\/} know something about the
system? Suppose we know the average of the outcomes,
\begin{equation}
\label{eq4-2}
\langle x\rangle =\frac{1}{N}\sum_{i=1}^N p_i x_i\;.
\end{equation}
Generalizing the principle of insufficient reason further, Jaynes
proposed that the probabilities $p_i$ in this case should be those
that maximize the information entropy (\ref{eq4-1}) given a fixed
value for $\langle x\rangle$, equation (\ref{eq4-2}).

%%%%%%%%%%%%%%%%%%%%%%%%%%%%%%%%%%%%%%%%%%
\begin{figure}[t]
\centerline{\hfill
\includegraphics[width=0.32\textwidth]{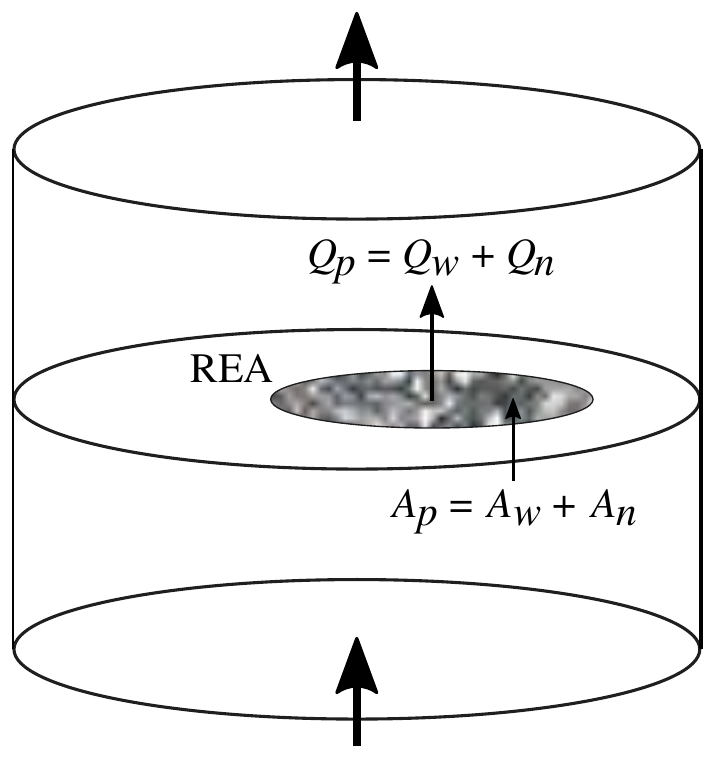}\hfill}
\caption{We define a plane that cuts through the cylindrical pore
  sample orthogonally to the average flow direction. We define
  Representative Elementary Area (REA) within the plane.  There is a
  wetting fluid flow rate $Q_w$ and a non-wetting fluid flow rate
  $Q_n$ passing through the REA. The total flow rate is $Q_p$. The
  wetting fluid covers an area $A_w$ of the REA and the non-wetting
  fluid an area $A_n$. The total pore area of the REA is
  $A_p$.\label{fig1}}
\end{figure}  
%%%%%%%%%%%%%%%%%%%%%%%%%%%%%%%%%%%%%%%%%%

Jaynes provided a set of circumstances for his generalized statistical
mechanics to be applicable. Hansen et al.\ \cite{hfss23} demonstrated
that these circumstances are met for steady-state immiscible two-phase
flow in porous media, and hence proceeded to construct such a
statistical mechanics.  The generalized statistical mechanics will
produce a thermodynamics-like framework at large scales.  This
thermodynamic-like framework is the end goal of this approach.

Suppose we have a cylindrical pore sample as shown in Figure
\ref{fig1}.  Two immiscible fluids enter at the bottom and leaves at
the top.  The side walls are close off. We assume that the pore
structure and chemical composition of the matrix to be statistically
uniform throughout the sample.  At some distance from the lower edge
of the sample, one may assume that the fluids have mixed sufficiently
to assume steady-state flow.  That is, if we make transversal cuts
through the sample, the structure of the porous medium and the
distribution of fluids in the pore space to be statistically uniform:
Comparing one cut to another, we will not be able to determine which
is closest to the inlet.  Figure \ref{fig1} shows one such cut.
Within this cut, we pick out a {\it Representative Elementary Area\/}
(REA), which is large enough to represent the statistics of the sample
and small enough to have the flow fluctuate.  There is an volumetric
flow rate $Q_p$ passing through the REA which may be split into a
volumetric flow rate of the wetting fluid $Q_w$ and the volumetric
flow rate of the non-wetting fluid $Q_n$, i.e.,
\begin{equation}
\label{eq2-1}
Q_p=Q_w+Q_n\;.
\end{equation}
The area of the REA is $A$. An area $A_p<A$ cuts through pore space.
This area may be divided into the area cutting through the pore space
filled with wetting fluid, $A_w$ and the area cutting though the pore
space filled with the non-wetting fluid, $A_n$, and we have that
\begin{equation}
\label{eq3-1}
A_p=A_w+A_n\;.
\end{equation}
At a given time for a given REA, the fluid configuration,
characterized by the velocity field and where the fluids are, is $X$.
The quantities that describe the REA depend on $X$ so that we have
$Q_p(X)$, $Q_w(X)$, $Q_n(X)=Q_p(X)-Q_w(X)$, $A_p(X)$, $A_w(X)$, and
$A_n(X)=A_p(X)-A_w(X)$.  Of course, $A_p(X)$ is not dependent on the
velocity field, nor the distribution of the two fluids, but only on
the shape of pore space. Let us denote $p(X)$ the probability density
to find fluid and pore space configuration $X$.  We define a {\it
  configurational entropy\/}\footnote{The probability density $p(X)$
is {\it continuous\/}, so that we are dealing with differential
entropy \cite{mnb14} in equation (\ref{eq4-3}). Another way of dealing
with the transition from discrete to continuous systems is that of
Jaynes \cite{j63}.}
\begin{equation}
\label{eq4-3}
S=-\int dX\ p(X)\ \ln p(X)\;.
\end{equation}
We assume that we know the averages (in time and position of the REA)
of the variables just described, i.e.,
\begin{eqnarray}
Q_u&=&\int dX\ p(X)\ Q_u(X)\;,\label{eq4-4}\\
A_w&=&\int dX\ p(X)\ A_w(X)\;,\label{eq4-5}\\
A_p&=&\int dX\ p(X)\ A_p(X)\;.\label{eq4-6}
\end{eqnarray}
The variable $Q_u$ is related to the volumetric flow rate $Q_p$, but
we will defer its definition until equation (\ref{eq4-32}).

Following the Jaynes maximum entropy principle, we maximize the
entropy (\ref{eq4-3}) given the constraints (\ref{eq4-4}) to
(\ref{eq4-6}), finding
\begin{equation}
\label{eq4-7}
p(X;\lambda_u,\lambda_w,\lambda_p)=\frac{1}{Z(\lambda_u,\lambda_w,\lambda_p)}\ \exp\left[-\lambda_u Q_u(X)-\lambda_w A_w(X)-\lambda_p A_p(X)\right]\;,
\end{equation}
where 
\begin{equation}
\label{eq4-8}
Z(\lambda_u,\lambda_w,\lambda_p)=\int\ dX\ \exp\left[-\lambda_u Q_u(X)-\lambda_w A_w(X)-\lambda_p A_p(X)\right]
\end{equation} 
is the normalization factor, also known as the partition
function. Three new {\em emergent variables\/} have appeared,
$\lambda_u$, $\lambda_w$, and $\lambda_p$.  We write the partition
function as
\begin{equation}
\label{eq4-13}
Z(\lambda_u,\lambda_w,\lambda_p)=\exp\left[-\lambda_u Q_z(\lambda_u,\lambda_w,\lambda_p)\right]\;.
\end{equation}

The partition function is in some sense the ``end product" of a
statistical mechanics approach to a problem.  It is the generating
function for the probability distribution for microscopic
configurations.  With it, all macroscopic averages may be calculated.

In ordinary statistical mechanics, a hamiltonian, i.e., a microscopic
energy function is necessary input for calculating the partition
function.  In the present problem we have three pore-level functions,
$Q_u(X)$, $A_w(X)$ and $A_p(X)$.  The problem in ordinary statistical
mechanics, is that even for seemingly simple problems such as the
two-dimensional Ising model for magnetic systems, calculating the
partition function may be extremely difficult, cfr.\ Onsager's
calculation of the partition function for this system \cite{o44}.  The
problem is surely no less difficult in the porous media
context. Perhaps, it is even worse in that we do not have good models
for the three microscopic functions in contrast to ordinary
statistical mechanics where model hamiltonians are legio.

A byproduct of the statistical mechanics approach is the emergence of
a set of thermo\-dynamics-like relations between the REA-scale
variables.  This framework will be in place whether it is possible to
calculate the partition function or not.  We will now develop this
thermodynamics-like framework for immiscible two-phase flow in porous
media.

We start by calculating the entropy $S$ by inserting equation
(\ref{eq4-7}) in equation (\ref{eq4-3}) and using equation
(\ref{eq4-13}).  We obtain
\begin{equation}
\label{eq4-13-1}
S(Q_z,A_w,A_p)=-\lambda_u Q_z(\lambda_u,\lambda_w,\lambda_p)+\lambda_u Q_u+\lambda_w A_w+\lambda_p A_p\;.
\end{equation}
We rewrite this equation as
\begin{equation}
\label{eq4-13-2}
Q_z=Q_u-\frac{1}{\lambda_u}\ S+\frac{\lambda_w}{\lambda_u}\ A_w +\frac{\lambda_p}{\lambda_u}\ A_p\;.
\end{equation}
It is convenient to define the new variables
\begin{eqnarray}
\theta&=&+\frac{1}{\lambda_u}\;,\label{eq4-15}\\
\mu&=&-\frac{\lambda_w}{\lambda_u}\;,\label{eq4-16}\\
\pi&=&-\frac{\lambda_p}{\lambda_u}\;.\label{eq4-17}
\end{eqnarray}
We may then write equation (\ref{eq4-13-2}) as
\begin{equation}
\label{eq4-13-3}
Q_u(S,A_w,A_p)=Q_z(\theta,\mu,\pi)+\theta S+\mu\ A_w +\pi\ A_p\;,
\end{equation}
where we have used that this is a Legendre transform\footnote{Assuming
convexity of the involved functions.} since
\begin{eqnarray}
S&=&-\left(\frac{\partial Q_z}{\partial \theta}\right)_{\mu,\pi}\;,\label{eq4-9}\\
A_w&=&-\left(\frac{\partial Q_z}{\partial \mu}\right)_{\theta,\pi}\;,\label{eq4-16}\\
A_p&=&-\left(\frac{\partial Q_z}{\partial \pi}\right)_{\theta,\mu}\;.\label{eq4-17}
\end{eqnarray}

We have finally arrived at defining the volumetric flow rate $Q_p$ in
equation (\ref{eq2-1}),
\begin{equation}
\label{eq4-31}
Q_p(\theta,A_w,A_p)=Q_z(\theta,\mu,\pi)+\mu A_w+\pi A_p\;,
\end{equation}
and the relation between $Q_u$ and $Q_p$ is then
\begin{equation}
\label{eq4-32}
Q_p(\theta,A_w,A_p)=Q_u(S,A_w,A_p)-S\theta\;.
\end{equation}

We see that $\theta$ (equation \ref{eq4-15}) plays the same role as
temperature in ordinary thermodynamics.  Hansen et al.\ \cite{hfss23}
named it {\it agiture,\/} which stands for {\it agitation
  temperature.\/} They also named $\mu$, (equation \ref{eq4-16}), the
{\it flow derivative.\/} It is analogous to {\it chemical potential\/}
in ordinary thermodynamics. Lastly, $\pi$, defined in equation
(\ref{eq4-17}) they called the {\it flow pressure.\/} It has no
equivalent in ordinary thermodynamics. We note that the unit of the
agiture is flow rate, whereas the unit of the flow derivative and the
flow pressure is velocity.

%%%%%%%%%%%%%%%%%%%%%%%%%%%%%%%%%%%%%%%%%%
\section{Extensive and intensive variables}
\label{intensive}

The variables $Q_u$, $Q_z$, $Q_p$, $A_w$, $A_p$, and $S$ are extensive
in the area of the REA: Double its area $A$, and all these variable
double. On the other hand, the variables $\theta$, $\mu$ and $\pi$ are
intensive.  They do not change when $A$ is changed.

We rescale $Q_p$, $A_w$ and $A_p$ by $A$, defining
\begin{eqnarray}
\phi&=&\frac{A_p}{A}\;,\label{eq5-1}\\
S_w\phi&=&\frac{A_w}{A}=\frac{A_w}{A_p}\phi\;,\label{eq5-2}\\
v_d&=&v_p\phi=\frac{Q_p}{A}=\frac{Q_p}{A_p}\phi\;,\label{eq5-3}\\
s&=&\frac{S}{A}\;,\label{eq5-6}
\end{eqnarray}
where $\phi$ is the porosity, $S_w$ the wetting saturation, $v_d$ the
Darcy velocity, $v_p$ the pore or seeping velocity, and $s$ the
entropy density.

In order to keep track of what happens to the variables in
e.g.\ $Q_p(\theta,A_w,A_p,A)$, we express the extensivity through the
scaling relation
\begin{equation}
\label{eq5-4}
\lambda Q_p(\theta,A_w,A_p,A)=Q_p(\theta,\lambda A_w,\lambda A_p,\lambda A)\;.
\end{equation}
We set the scale factor $\lambda=1/A$, finding
\begin{equation}
\label{eq5-5}
v_d(\theta,S_w,\phi)=v_p(\theta,S_w,\phi)\phi=\frac{1}{A}Q_p(\theta,A_w,A_p,A)=Q_p(\theta,S_w\phi,\phi,1)\;.
\end{equation}
Using the same arguments we find that
\begin{equation}
\label{eq5-7}
s=s(\theta,S_w,\phi)\;.
\end{equation}

%%%%%%%%%%%%%%%%%%%%%%%%%%%%%%%%%%%%%%%%%%
\begin{figure}[t]
\centerline{\hfill
\includegraphics[width=0.18\textwidth]{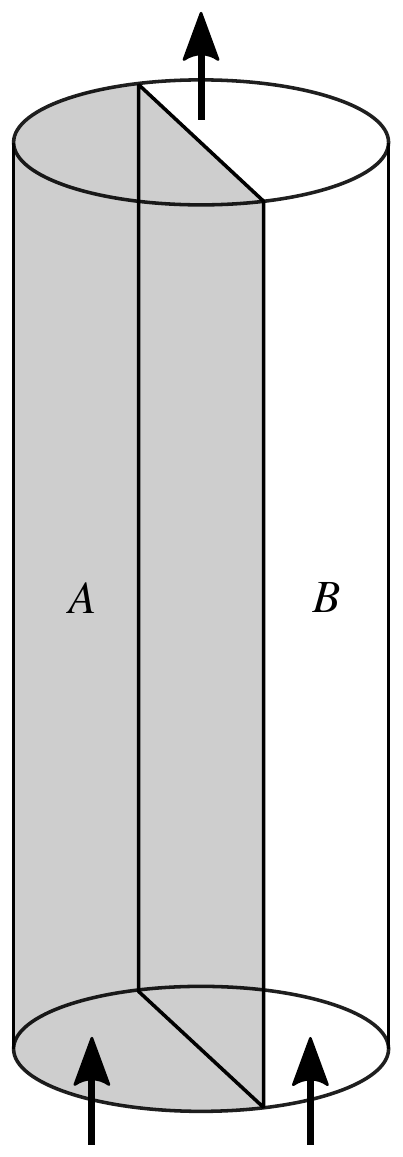}\hfill}
\caption{A cylindrical porous media sample consisting of two halves A
  and B having different properties with respect to the two immiscible
  fluids injected at the bottom edge. The two halves are in direct
  contact and fluids may pass unrestricted between the two halves.
\label{fig2}}
\end{figure}  
%%%%%%%%%%%%%%%%%%%%%%%%%%%%%%%%%%%%%%%%%%

%%%%%%%%%%%%%%%%%%%%%%%%%%%%%%%%%%%%%%%%%%
\section{Emergent variables}
\label{emergent}

The three variables $\theta$, $\mu$ and $\pi$ emerge as a result of
the scale-up process.  They are the conjugate variables to the entropy
$S$, the wetting area $A_w$ and the pore area $A_p$ respectively.  How
are we to interpret them?

We follow Hansen et al.\ \cite{hfss23} and consider a cylindrical
porous medium, see figure \ref{fig2}.  We imagine a plane running
through the cylinder axis parallel to the average flow direction.  On
one side of the plane, the porous medium has one set of properties, on
the other side of the plane it has another set of properties.  The
difference may be in chemical composition of the matrix or it may
consist in different porosity or topological structure.  We name the
two halves ``A" and ``B" respectively.  We assume the two halves to be
statistically homogeneous along the axis.

We now assume that two immiscible and incompressible fluids are
simultaneously pushed through the porous cylinder.  Away from the edge
where the two fluids are injected, the flow is in a steady state.
This means that the flow statistics is invariant along the axis.
However, it does not mean that the fluid-fluid interfaces do not move.
At sufficient flow rates, they do and as a result, fluid clusters will
merge and break up.

As the statistical distributions describing the flow is invariant
along the flow axis, the configurational entropy $S$ is invariant
along this axis.  As shown by Hansen et al.\ \cite{hfss23}, this means
that the agiture in sectors A and B, $\theta^A$ and $\theta^B$ are
equal,
\begin{equation}
\label{eq6-1}
\theta^A=\theta^B\;.
\end{equation}

In ordinary thermodynamics, the rule is that the conjugate of a
conserved quantity is constant in a heterogeneous system at
equilibrium. This is a generalization of the argument for the
temperature being the same everywhere in a system at equilibrium as
the entropy is conserved and the temperature is its conjugate. This
reasoning may be repeated for the flow problem.

Neither $A_w$ nor $A_p$ are conserved along the flow axis at the pore
scale.  This would imply that $\mu^A\neq \mu^B$ and $\pi^A\neq \pi^B$.
This remains true at the continuum scale for $A_p$.  However, $A_w$ is
conserved at the continuum scale, see equation (\ref{eq1-7}).  This is
an emergent conservation law. We therefore conjecture that
\begin{equation}
\label{eq6-2}
\mu^A=\mu^B\;.
\end{equation}

%%%%%%%%%%%%%%%%%%%%%%%%%%%%%%%%%%%%%%%%%%
\subsection{Agiture}
\label{agiture}

The agiture $\theta$ plays the role of a temperature in the
thermodynamics-like formalism that we are developing.  But, can we
express it in terms of more familiar variables?

We note first that in the formalism we have developed in Sections
\ref{statmech} and \ref{intensive}, there is one variable that is
missing: the pressure $P$. Intuitively, one would expect the agiture
to be related to the pressure gradient $\nabla P$: Higher pressure
gradient should indicate higher agiture.

%%%%%%%%%%%%%%%%%%%%%%%%%%%%%%%%%%%%%%%%%%
\begin{figure}[b]
\centerline{\hfill
\includegraphics[width=0.18\textwidth]{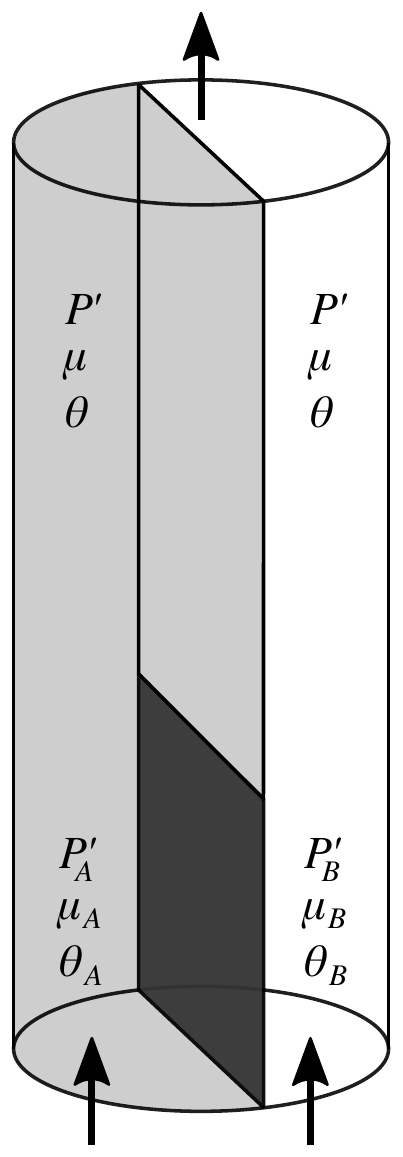}\hfill}
\caption{A cylindrical porous media sample consisting of two halves A
  and B having different properties with respect to the two immiscible
  fluids injected at the bottom edge. There is a impenetrable wall
  separating the two halves at the bottom third of the boundary. This
  is indicated by the dark section. Above, the two halves are in
  direct contact as in \ref{fig2}. A mixture of the two fluids are
  injected into the two halves at the lower edge. The pressure
  gradient is $P'_A$ on the A side and $P'_B$ on the B side up to the
  upper edge of the impenetrable wall. Likewise the flow derivatives
  are $\mu_A$ and $\mu_B$ and the agitures are $\theta_A$ and
  $\theta_B$.  Higher up where the two halves communicate, the
  pressure gradients, the flow derivatives and the agitures become
  pairwise equal.\label{fig3}}
\end{figure}  
%%%%%%%%%%%%%%%%%%%%%%%%%%%%%%%%%%%%%%%%%%

Figure \ref{fig3} shows a cylindrical porous medium consisting of two
halves A and B.  They have different properties such as porosity,
chemical composition of the matrix, or the pore spaces have different
topology. At the bottom one third of the porous cylinder, the boundary
between the two halves is impenetrable. This is indicated by the dark
section of the boundary.  Above, the two halves are in direct contact.
We inject the immiscible fluids into each of the halves A and B at the
bottom.  We may do this by injecting at different volumetric flow
rates $Q_p^A=Q_w^A+Q_n^A$ and $Q_p^B=Q_w^B+Q_n^B$.  This gives rise to
flow derivatives $\mu_A$ and $\mu_B$, and agitures $\theta_A$ and
$\theta_B$ in the two halves far enough from the inlet for the flow to
be in a steady state.  The pressure gradient will then point along the
axis of the cylinder, the $z$-direction, and we will in the following
use the notation $P'=\partial P/\partial z$ for the gradient. We find
a pressure gradient $P'_A$ and $P'_B$ in A and B respectively. Further
into the cylindrical sample above the impenetrable wall, the flow will
adjust to a new steady state characterized by equal agiture $\theta$
and flow derivative $\mu$ and in both halves according to equations
(\ref{eq6-1}) and (\ref{eq6-2}).  Likewise the pressure gradient $P'$
must be equal in the two halves since there cannot be any net flow in
either direction across the boundary between the two halves, apart
from local fluctuations.

Since the two variables $\theta$ and $\mu$ on one side and $P'$ on the
other side form alternate descriptions, we must assume that they are
related.  The simplest relation between them we may write down is
\begin{equation}
\label{eq6-4}
P' = - c_\theta \theta +c_\mu \mu\;.
\end{equation}
The reason for the minus sign in front of the $\theta$ term is due to
the flow direction is the opposite of the pressure gradient
direction. We note that the unit of $c_\theta$ is viscosity and that
of $c_\mu$ is viscosity times area.

We note that equation (\ref{eq5-5}) defines the seepage velocity
$v_p(\theta,S_w,\phi)$ where $S_w$ rather than the flow derivative
$\mu$ is a control variable.  At the same time, the seepage velocity
depends on the local pressure gradient, $v_p(P',S_w,\phi)$.  Hence, we
conjecture that $c_\mu=0$ so that
\begin{equation}
\label{eq6-5}
P' = - c_\theta \theta\;.
\end{equation}
We note that $Q_p=Q_p(\theta,A_w,A_p)$ defined in equation
(\ref{eq4-31}) may be written
\begin{equation}
\label{eq6-6}
Q_p=Q_p(P',A_w,A_p)=A_p v_p(P',S_w,\phi)\;,
\end{equation}
where we also have used equation (\ref{eq5-5}).  We see that this is a
constitutive equation relating seepage velocity to local pressure
gradient, saturation and porosity.

To be concrete, we return for a moment to Section \ref{relperm} and
the constitutive equations forming the core of relative permeability
theory, (\ref{eq1-1}) and (\ref{eq1-2}), which give
\begin{equation}
\label{eq6-7}
v_p(P',S_w,\phi)=-\frac{K}{\phi}\left[\frac{k_{rw}(S_w)}{\mu_w}+\frac{k_{rn}(S_w)}{\mu_n}\right] P'\;.
\end{equation}

The agiture is the conjugate variable to the entropy.  We may express
it as
\begin{equation}
\label{eq6-8}
\theta=\left(\frac{\partial Q_u}{\partial S}\right)_{A_w,A_p}\;,
\end{equation}
The entropy is given by 
\begin{equation}
\label{eq6-9}
S=-\left(\frac{\partial Q_p}{\partial \theta}\right)_{A_w,A_p}=c_\theta\left(\frac{\partial Q_p}{\partial P'}\right)_{A_w,A_p}\;,
\end{equation} 
or in terms of the entropy density,
\begin{equation}
\label{eq6-10}
s=c_\theta\left(\frac{\partial v_p}{\partial P'}\right)_{S_w,\phi}\;.
\end{equation} 

We note that the {\it differential mobility\/} is defined as 
\begin{equation}
\label{eq6-11}
m(P',S_w,\phi)=-\left(\frac{\partial v_p}{\partial P'}\right)_{S_w,\phi}\;.
\end{equation}
Using the relative permeability constitutive equation (\ref{eq6-7}),
we find the mobility\footnote{Mobility and differential mobility are
equal in this case as the velocity is linear in the pressure
gradient.} in this case to be
\begin{equation}
\label{eq6-12}
m(S_w,\phi)=\frac{K}{\phi}\left[\frac{k_{rw}(S_w)}{\mu_w}+\frac{k_{rn}(S_w)}{\mu_n}\right]\;.
\end{equation}

Comparing equations (\ref{eq6-10}) and (\ref{eq6-11}), we see
that\footnote{Since we are dealing with {\it differential entropy,\/}
it is not a problem that it is negative.}
\begin{equation}
\label{eq6-13}
s=-c_\theta m\;.
\end{equation}
It is surprising that the differential mobility and the entropy should
be related.

Equation (\ref{eq6-5}) relates the agiture $\theta$ and the pressure
gradient $P'$ and the way we have argued is by demonstrating that the
pressure gradient behaves as one would expect an agiture (= agitation
temperature) should do. However, could one test equation (\ref{eq6-5})
directly by measuring the left hand side and compare it to the right
hand side. This is difficult as it would entail measuring the
configurational entropy and then use equation (\ref{eq6-8}); and
measuring entropy is difficult, also in ordinary thermodynamics.

%%%%%%%%%%%%%%%%%%%%%%%%%%%%%%%%%%%%%%%%%%
\subsection{Flow derivative}
\label{mu}

Hansen et al.\ \cite{hsbkgv18} defined the two {\it thermodynamic velocities,\/}
\begin{eqnarray}
{\hat v}_w=\left(\frac{\partial Q_p}{\partial A_w}\right)_{\theta,A_n}\;,\label{eq6-14}\\
{\hat v}_n=\left(\frac{\partial Q_p}{\partial A_n}\right)_{\theta,A_w}\;,\label{eq6-15}
\end{eqnarray}
where the control variables are $\theta$, $A_w$ and $A_n$ defined in
(\ref{eq3-1}), making $A_p$ a dependent variable. Changing the
variables $(\theta,A_w,A_n)\to(\theta,S_w,A_p)$, these two equations
may be written
\begin{eqnarray}
{\hat v}_w=v_p+S_n\left(\frac{\partial v_p}{\partial S_w}\right)_{\theta,\phi}\;,\label{eq6-16}\\
{\hat v}_n=v_p-S_w\left(\frac{\partial v_p}{\partial S_w}\right)_{\theta,\phi}\;.\label{eq6-17}
\end{eqnarray}
The flow derivative $\mu$, which is the conjugate of the wetting area
$A_w$ and thereby also the saturation $S_w$, is given by
\begin{equation}
\label{eq6-18}
\mu=-\left(\frac{\partial Q_p}{\partial A_w}\right)_{\theta,A_p}=\left(\frac{\partial v_p}{\partial S_w}\right)_{\theta,\phi}\;.
\end{equation}
Hence, equation (\ref{eq6-17}) may be recognized as a Legendre
transformation substituting $S_w \to \mu$:
\begin{equation}
\label{eq6-19}
{\hat v}_n(\theta,\mu,\phi)=v_p(\theta,S_w,\phi)-S_w(\theta,\mu,\phi)\mu\;.
\end{equation}
In other words, the non-wetting thermodynamic velocity is the Legendre
transformation of the average seepage velocity with respect to the
saturation.

%%%%%%%%%%%%%%%%%%%%%%%%%%%%%%%%%%%%%%%%%%
\subsection{Flow pressure}
\label{pi}

The flow pressure $\pi$ is the conjugate of the pore area $A_p$, and
thereby the porosity $\phi$,
\begin{equation}
\label{eq6-20}
\pi=-\left(\frac{\partial Q_p}{\partial A_p}\right)_{\theta,A_w}=\left(\frac{\partial v_p}{\partial \phi}\right)_{\theta,S_w}\;.
\end{equation}
There are no conservation laws associated with this variable. 

We expect the flow pressure $\pi$ to be of less practical use, as we
expect that the flow rate $Q_p$ to be proportional to the pore area
$A_p$. This makes the Legendre transform between $A_p$ and $\pi$ break
down due to lack of convexity. The flow pressure will nevertheless
play a role e.g., when porosity gradients are present.

%%%%%%%%%%%%%%%%%%%%%%%%%%%%%%%%%%%%%%%%%%
\section{Seepage velocities and the co-moving velocity}
\label{seepage}

The thermodynamic velocities defined in equations (\ref{eq6-14}) and
(\ref{eq6-15}) are not the seepage velocities which we define as
\begin{eqnarray}
{v}_w=\frac{Q_w}{A_w}\;,\label{eq7-1}\\
{v}_n=\frac{Q_n}{A_n}\;.\label{eq7-2}
\end{eqnarray}
These are the velocities that measured in the laboratory, whereas the
thermodynamic velocities are not. We rewrite equation (\ref{eq2-1}) as
\begin{equation}
\label{eq7-10}
Q_p=Q_w+Q_n=v_wA_w+v_nA_n\;.
\end{equation}
We express the saturations in terms of the variables $A_w$ and $A_n$,
\begin{equation}
\label{eq7-12}
S_w=\frac{A_w}{A_w+A_n}\;,
\end{equation}
and
\begin{equation}
\label{eq7-13}
S_n=\frac{A_n}{A_w+A_n}\;,
\end{equation}
leading to equation (\ref{eq1-3}). We may then write equation
(\ref{eq7-10}) as
\begin{equation}
\label{eq7-15}
v_p=v_wS_w+v_nS_n\;,
\end{equation}
where $v_p$, the average velocity, is defined in equation
(\ref{eq5-5}).

We now turn to the thermodynamic velocities, equations (\ref{eq6-14})
and (\ref{eq6-15}). The average flow rate $Q_p$ obeys the scaling
relation Euler scaling
\begin{equation}
\label{eq7-16}
\lambda Q_p(P',A_w,A_n)=Q_p(P',\lambda A_w,\lambda A_n)\;.
\end{equation}
We then have from the Euler theorem for homogeneous functions
\begin{equation}
\label{eq7-17}
Q_p=\left(\frac{\partial Q_p}{\partial A_w}\right)_{A_n,\phi,P'}A_w+\left(\frac{\partial Q_p}{\partial A_n}\right)_{A_w,\phi,P'}A_n=\hat{v}_wA_w+\hat{v}_nA_n\;.
\end{equation}
Dividing this equation by $A_p$ gives
\begin{equation}
\label{eq7-18}
v_p={\hat v}_w S_w+{\hat v}_n S_n\;.
\end{equation}

Comparing equations (\ref{eq7-15}) and (\ref{eq7-18}),
\begin{equation}
\label{eq7-19}
v_p=v_wS_w+v_nS_n={\hat v}_w S_w+{\hat v}_n S_n\;.
\end{equation}
This equation does not imply that $v_w={\hat v}_w$ and $v_n={\hat
  v}_n$.  Rather, the most general relation between the thermodynamic
and seeping velocities are
\begin{eqnarray}
v_w={\hat v}_w-S_n v_m\;,\label{eq7-20}\\
v_n={\hat v}_n+S_w v_m\;,\label{eq7-21}
\end{eqnarray}
where $v_m$ is the {\it co-moving velocity\/}
\cite{hsbkgv18,rsh20,ph23}.  We may combine these two equations with
equatons (\ref{eq6-16}) and (\ref{eq6-17}) to find
\begin{eqnarray}
v_w&=&v_p+S_n\left[\left(\frac{\partial v_p}{\partial S_w}\right)_{\phi,P'}-v_m\right]\;,\label{eq7-22}\\
v_n&=&v_p-S_w\left[\left(\frac{\partial v_p}{\partial S_w}\right)_{\phi,P'}-v_m\right]\;.\label{eq7-23}
\end{eqnarray}
By taking the derivative of equation (\ref{eq7-19}) with respect to
$S_w$ and using equations (\ref{eq7-20}) and (\ref{eq7-21}), we find
\begin{equation}
\label{eq2-23} 
v_m=\left(\frac{\partial v_p}{\partial S_w}\right)_{\phi,P'}-v_w+v_n=S_w\left(\frac{\partial v_w}{\partial S_w}\right)_{\phi,P'}+S_n\left(\frac{\partial v_n}{\partial S_w}\right)_{\phi,P'}\;.
\end{equation}

We now return to equation (\ref{eq6-19}), which may now be expressed
in terms of the seepage velocity $v_n$ rather than the thermodynamic
velocity ${\hat v}_n$,
\begin{equation}
\label{eq7-24}
v_n(\theta,\mu,\phi)-S_w(\theta,\mu,\phi)v_m(\theta,\mu,\phi)=v_p(\theta,S_w,\phi)-S_w(\theta,\mu,\phi)\mu\;,
\end{equation}
signifying that $(\theta,\mu,\phi)$ are the natural variables for the
co-moving velocity. Measurements of the co-moving velocity suggest
that it has the simple functional form \cite{rpsh22,amphbma24,h24}
\begin{equation}
\label{eq7-25}
v_m(\theta,\mu,\phi)=a(\theta,\phi)+b(\theta,\phi)\mu\;,
\end{equation}
based on analysis of relative permeability data, on dynamic pore
network modeling and on lattice Boltzmann simulations on reconstructed
sandstones.

It has recently been realized that the co-moving velocity has an
equivalent in the thermodynamics of two-fluid mixtures \cite{ohhl25},
where a {\it co-molar volume\/} ties the partial molar volumes of a
binary mixture together with the Voronoi volumes of each of the fluids
in the mixture.  When plotting the co-molar volume against the
variable corresponding to $\mu$, it is {\it nearly\/} linear, but not
quite.

We note that the thermodynamic velocities $({\hat v}_w,{\hat v}_n)$
can be found by knowing $v_p$ alone, see equations (\ref{eq6-16}) and
(\ref{eq6-17}).  The seepage velocities, on the other hand requires
additional knowledge, i.e., $v_p$ and $v_m$, and are given by
equations (\ref{eq7-22}) and (\ref{eq7-23}). They provide a two-way
mapping
\begin{equation}
\label{eq7-26}
\begin{pmatrix}
v_w\\
v_n\\
\end{pmatrix}
\leftrightarrow
\begin{pmatrix}
v_p\\
v_m\\
\end{pmatrix}
\;.
\end{equation}

%%%%%%%%%%%%%%%%%%%%%%%%%%%%%%%%%%%%%%%%%%
\section{Control variables}
\label{control}

In Section \ref{statmech} we made a choice when focusing on the
averages $Q_u$, $A_w$ and $A_p$ in equations (\ref{eq4-4}) to
(\ref{eq4-6}), leading to a thermodynamics-like formalism based upon
the extensive variables $Q_p$, $A_w$, and $A_p$ and their intensive
conjugates $\theta$ ($=-c_\theta P'$), $\mu$, and $\pi$.  In core
flooding, these are not the natural control variables.  Rather, one
controls the flow rates $Q_w$ and $Q_n$ into the core, making $A_w$
(i.e.\ the saturation) and the pressure gradient the dependent
variables.  It is possible to build a Jaynes statistical mechanics
based on $Q_u$, $Q_w$ and $A_p$ rather than $Q_u$, $A_w$ and $A_p$.
We will do this elsewhere.  However, we can still develop the ensuing
thermodynamics-like formalism as long as we work only with the
extensive variables.  We will do so in the following.

%%%%%%%%%%%%%%%%%%%%%%%%%%%%%%%%%%%%%%%%%%
\subsection{Controlling the fractional flow rate}
\label{flowrate}

Our starting point is again the REA shown in figure \ref{fig1}, and
equations (\ref{eq2-1}) and (\ref{eq3-1}),
\begin{equation}
\tag{\ref{eq2-1}}
Q_p=Q_w+Q_n\;,
\end{equation}
and
\begin{equation}
\tag{\ref{eq3-1}}
A_p=A_w+A_n\;.\nonumber
\end{equation}
We now assume that our control variables are $Q_w$ and $Q_n$ together
with $A_p$, the REA pore space. The dependent variable is the pressure
gradient $P'$.  They are related through the constitutive equation
\begin{equation}
\label{eq8-1}
P'=P'(Q_w,Q_n,A_p)\;.
\end{equation}
We may rescale the extensive variables in this expression,
$A_p\to\lambda A_p$, $Q_w\to\lambda Q_w$, and $Q_n\to\lambda
Q_n$.\footnote{In practice, changing $A_p$ means changing the
sample. Hence, this is a theoretical exercise. Nevertheless,
mathematics works.}  This leaves $P'$ unchanged, i.e.,
\begin{equation}
\label{eq8-1-1}
P'(\lambda Q_w,\lambda Q_n,\lambda A_p)=P'(Q_w,Q_n,A_p)\;.
\end{equation}

It is, however, convenient to rewrite equation (\ref{eq8-1}) as 
\begin{equation}
\label{eq8-2}
A_p=A_p(P',Q_w,Q_n)\;,
\end{equation}
making it implicit with respect to $P'$. The scaling relation
(\ref{eq8-1-1}) then takes the form
\begin{equation}
\label{eq8-3}
\lambda A_p(P',Q_w,Q_n)=A_p(P',\lambda Q_w,\lambda Q_n)\;.
\end{equation}
 
We then use the Euler theorem for homogeneous functions,
\begin{equation}
\label{eq8-4}
A_p=\left(\frac{\partial A_p}{\partial Q_w}\right)_{Q_n,\phi,P'}Q_w+\left(\frac{\partial A_p}{\partial Q_n}\right)_{Q_w,\phi,P'}Q_n
=\hat{p}_wQ_w+\hat{p}_nQ_n\;,
\end{equation}
where we have defined the {\it thermodynamic paces\/}\footnote{Pace is
inverse velocity.}
\begin{equation}
\label{eq8-5}
\hat{p}_w=\left(\frac{\partial A_p}{\partial Q_w}\right)_{Q_n,\phi,P'}\;,
\end{equation}
and
\begin{equation}
\label{eq8-6}
\hat{p}_n=\left(\frac{\partial A_p}{\partial Q_n}\right)_{Q_w,\phi,P'}\;.
\end{equation}

We also define the {\it physical paces\/}
\begin{equation}
\label{eq8-7}
p_p=\frac{A_p}{Q_p}=\frac{1}{v_p}\;,
\end{equation}
\begin{equation}
\label{eq8-8}
p_w=\frac{A_w}{Q_w}=\frac{1}{v_w}\;,
\end{equation}
and
\begin{equation}
\label{eq8-9}
p_n=\frac{A_n}{Q_n}=\frac{1}{v_n}\;.
\end{equation}
This allows us to write equation (\ref{eq3-1}) as
\begin{equation}
\label{eq8-10}
A_p=p_wQ_w+p_nQ_n\;.
\end{equation}
Combining this equation with equation (\ref{eq8-4}) gives
\begin{equation}
\label{eq8-11}
A_p=p_wQ_w+p_nQ_n=\hat{p}_wQ_w+\hat{p}_nQ_n\;.
\end{equation}

We now introduce the fractional flow rates
\begin{equation}
\label{eq8-12}
F_w=\frac{Q_w}{Q_w+Q_n}\;,
\end{equation}
and
\begin{equation}
\label{eq8-13}
F_n=\frac{Q_n}{Q_w+Q_n}\;,
\end{equation}
so that 
\begin{equation}
\label{eq8-14}
F_w+F_n=1\;.
\end{equation}
We may then write equation (\ref{eq8-11}) as
\begin{equation}
\label{eq8-15}
p_p=p_wF_w+p_nF_n=\hat{p}_wF_w+\hat{p}_nF_n\;,
\end{equation}
where $p_p$, the average pace, is defined in equation (\ref{eq8-7}).

We now make a coordinate transformation, 
\begin{equation}
\label{eq8-16}
(P',Q_w,Q_n)\to (P',Q_p,F_w)\;,
\end{equation}
so that we have 
\begin{equation}
\label{eq8-17}
A_p(P',Q_p,F_w)=A_p(P',Q_w,Q_n)\;.
\end{equation}
We use extensivity,
\begin{equation}
\label{eq8-18}
\lambda A_p(P',Q_p,F_w)=A_p(P',\lambda Q_p,F_w)\;,
\end{equation}
leading to
\begin{equation}
\label{eq8-19}
A_p(P',Q_p,F_w)=Q_pA_p(P',1,F_w)=Q_p p_p(P',F_w)\;,
\end{equation}
when setting $\lambda=1/Q_p$ and defining $p_p(P',F_w)=A_p(P',1,F_w)$.  

We express the thermodynamic paces equations (\ref{eq8-5}) and
(\ref{eq8-6}) in terms of $(P',Q_p,F_w)$, finding
\begin{equation}
\label{eq8-20}
\hat{p}_w=p_p+F_n\left(\frac{\partial p_p}{\partial F_w}\right)_{\phi,P'}\;,
\end{equation}
and 
\begin{equation}
\label{eq8-21}
\hat{p}_n=p_p-F_w\left(\frac{\partial p_p}{\partial F_w}\right)_{\phi,P'}\;.
\end{equation}
We may define a {\it co-moving pace function,\/} $p_m$ by the expressions
\begin{equation}
\label{eq8-22}
\hat{p}_w=p_w+F_n p_m\;,
\end{equation}
and
\begin{equation}
\label{eq8-23}
\hat{p}_n=p_n-F_w p_m\;,
\end{equation}
where $p_m$ is the most general function to fulfill equation
(\ref{eq8-15}). Combining these two equations with equations
(\ref{eq8-21}) and (\ref{eq8-22}) gives
\begin{eqnarray}
p_w&=&p_p+F_n\left[\left(\frac{\partial p_p}{\partial F_w}\right)_{\phi,P'}-p_m\right]\;,\label{eq8-24}\\
p_n&=&p_p-F_w\left[\left(\frac{\partial p_p}{\partial F_w}\right)_{\phi,P'}-p_m\right]\;,\label{eq8-25}
\end{eqnarray}
By taking the derivative of equation (\ref{eq8-15}) with respect to
$F_w$ and using equations (\ref{eq8-22}) and (\ref{eq8-23}), we find
\begin{equation}
\label{eq8-26} 
p_m=\left(\frac{\partial p_p}{\partial F_w}\right)_{\phi,P'}-p_w+p_n=F_w\left(\frac{\partial p_w}{\partial F_w}\right)_{\phi,P'}+F_n\left(\frac{\partial p_n}{\partial F_w}\right)_{\phi,P'}\;.
\end{equation}

\begin{figure}[t]
\centerline{\hfill
\includegraphics[width=0.6\textwidth]{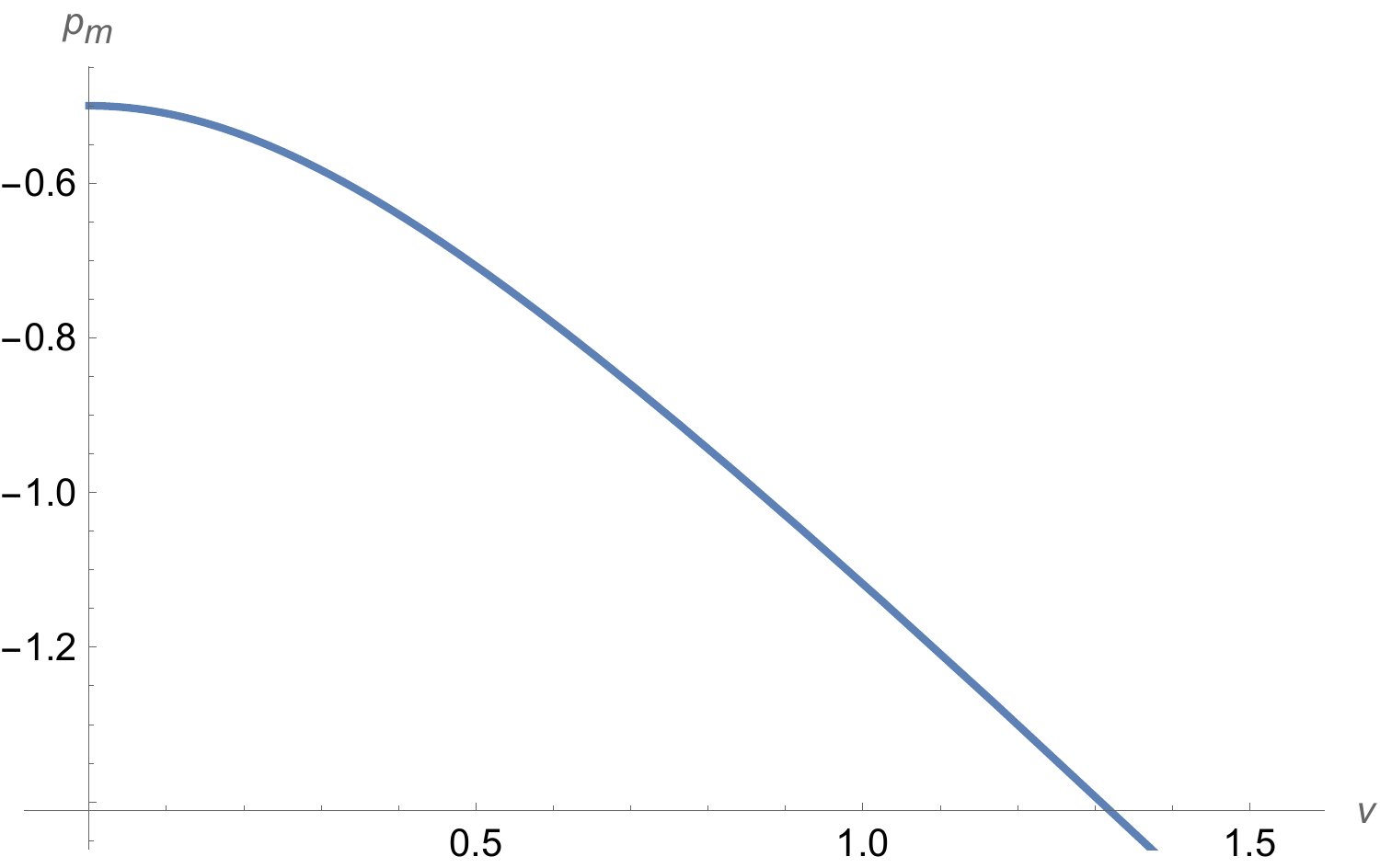}\hfill}
\caption{The co-moving pace $p_m$ against $\nu$ for the Picchi and
  Battiato pair of relative permeabilities \cite{pb19}, equations
  (\ref{eq8-27}) and (\ref{eq8-28}) with $\mu_w=\mu_n$,
  $k_{rw}^0=k_{rn}^0/2$, and $\alpha=1/2$.
\label{fig4}}
\end{figure}

We may at this point ask whether the co-moving pace $p_m$ is a linear
function in the variable
\begin{equation}
\label{eq8-26-1}
\nu=\left(\frac{\partial p_p}{\partial F_w}\right)_{\phi,P'}
\end{equation} 
in the same way that the co-moving velocity $v_m$ is a linear function
of $\mu=(\partial v_p/\partial S_w)_{\phi,P'}$, see equation
(\ref{eq7-25}).  The answer is no.  This may e.g., be seen by testing
the relative permeability pair proposed by Picchi and Battiato
\cite{pb19}
\begin{eqnarray}
k_{rw}&=&k_{rw}^0 S_w^2\;,\label{eq8-27}\\
k_{rn}&=&k_{rn}^0 S_n^2\left[1+2\frac{\alpha\mu_n}{\mu_w}\frac{S_w}{S_n}\right]\;,\label{eq8-28}
\end{eqnarray}
where $0\le \alpha \le 1$ is a parameter.  This pair leads to the co-moving velocity
\begin{equation}
\label{eq8-29}
v_m=\alpha k_{rn}^0 +\frac{\mu}{2}\;,
\end{equation}
in units of $v_0=-KP'/\mu_w\phi$. The corresponding co-moving pace is
a complex curve when plotting $p_m$ vs.\ $\nu$, see figure \ref{fig4}.

%%%%%%%%%%%%%%%%%%%%%%%%%%%%%%%%%%%%%%%%%%
%%%%%%%%%%%%%%%%%%%%%%%%%%%%%%%%%%%%%%%%%%

We note that we used the following relations between fractional flow
$F_w$ and saturation $S_w$ in calculating $p_m$ above:
\begin{equation}
\label{eq8-27}
F_w=\frac{Q_w}{Q_p}=\frac{A_p S_w v_w}{A_p v_p}=S_w\ \frac{v_w(P',S_w,\phi)}{v_p(P',S_w,\phi)}\;.
\end{equation}
Likewise, we have
\begin{equation}
\label{eq8-28}
S_w=\frac{A_w}{A_p}=\frac{Q_p F_w p_w}{Q_p p_p}=F_w\ \frac{p_w(P',F_w,\phi)}{p_p(P',F_w,\phi)}\;.
\end{equation}

%%%%%%%%%%%%%%%%%%%%%%%%%%%%%%%%%%%%%%%%%%
\section{Discussion and conclusion}
\label{discussion}

The aim of this review has been to present the central ideas of the
thermodynamics-like formalism for immiscible two-phase flow in porous
media that was first introduced by Hansen et al.\ \cite{hsbkgv18} in
the context of steady-state flow.  The ultimate goal, however, of this
approach is to be able to handle flow that is not in a steady state.
This means handling gradients in the conjugate variables to
configurational entropy, saturation and porosity: agiture $\theta$,
flow derivative $\mu$ and flow pressure $\pi$, turning the equilibrium
thermodynamics-like formalism into a non-equilibrium
thermodynamics-like formalism.

We noted in Section \ref{statmech} that actually calculating the
partition function, equation (\ref{eq4-13}), may be impossible in the
present context. The consequence of this is that we are unable to
derive directly the constitutive equation $Q_p(P',A_w,A_p)$ from the
pore level physics. It must be found by other means, and then used as
input in the thermodynamics-like formalism.

We have conjectured here that the agiture, the temperature-like
emergent variable conjugate to the configurational entropy, is
proportional to the pressure gradient.  A surprising consequence of
this conjecture is that the differential mobility of the fluids is
directly related to the entropy: higher differential mobility means
lower configurational entropy.

The question of what are the control variables and the dependent
variables is central, and much of the thermodynamics-like formalism
concerns how to switch between the extensive variables $Q_p$, $A_w$,
and $A_p$, and their intensive conjugates, $\theta$, $\mu$, and $\pi$.
However, there was already a choice made when the three extensive
variables were picked. We could have made other choices.  We discuss
for the first time in this review another choice: $Q_p$, $Q_w$ and
$A_p$.  Without developing a Jaynes-type statistical mechanics based
on this choice, we nevertheless explore aspects of the ensuing
thermodynamics-like formalism.  We find a set of equations that
parallel those found in \cite{hsbkgv18}, but where the fractional flow
rate $F_w$ rather than the saturation $S_w$ is a control
variable. Since this is the natural variable in core flooding
experiments, not the saturation, such a formalism is necessary.

%%%%%%%%%%%%%%%%%%%%%%%%%%%%%%%%%%%%%%%%%%
\bigskip

\noindent
We thank Steffen Berg for valuable discussions and comments.

\section*{\normalsize Author contributions}
The authors have contributed equally to all aspects of this work.

\section*{\normalsize Funding}
This work was partly supported by the Research Council of
  Norway through its Centers of Excellence funding scheme, project
  number 262644. We furthermore acknowledge funding from the European
  Research Council (Grant Agreement 101141323 AGIPORE).

\xdash[6em]
\end{document}